\documentclass[12pt]{iopart}
\usepackage[dvipdfmx]{graphicx}
\newcommand{\sr}{SrPt$_2$Sb$_2$}
\newcommand{\sras}{SrPt$_2$As$_2$}
\newcommand{\cabe}{CaBe$_2$Ge$_2$}
\newcommand{\tz}{\ensuremath{T_{\rm 0}}}
\newcommand{\tc}{\ensuremath{T_{\rm c}}}

\begin{document}

\title{Possibility of charge density wave transition in {\sr} superconductor}

\author{Soshi~Ibuka\footnote{Present address: High Energy Accelerator Research Organization, Tokai, Ibaraki 319-1106, Japan.} and Motoharu~Imai}
\address{National Institute for Materials Science, Tsukuba, Ibaraki 305-0047, Japan}
\ead{ibuka@post.j-parc.jp}

\begin{abstract}
The first-order transition at ${\tz} = 270$~K for the platinum-based {\sr} superconductor was investigated using X-ray diffraction and magnetic susceptibility measurements. When polycrystalline {\sr} was cooled down through {\tz}, the structure was transformed from monoclinic to a modulated orthorhombic structure, and no magnetic order was formed, which illustrates the possibility of a charge density wave (CDW) transition at {\tz}. {\sr} can thus be a new example to examine the interplay of CDW and superconductivity in addition to SrPt$_2$As$_2$, BaPt$_2$As$_2$, and LaPt$_2$Si$_2$. It is unique that the average structure of the low-temperature phase has higher symmetry than that of the high-temperature phase.
\end{abstract}


\vspace{2pc}

\submitto{\JPCM}

\maketitle

\section{Introduction}
After the coexistence of a charge density wave (CDW) and superconductivity was identified in quasi-two-dimensional transition metal dichalcogenides~\cite{GabovichAM2001}, there has been an increasing interest in the mechanism of CDW and the relation to superconductivity. CDW is a modulation of the conduction electron density associated with a modulation of the atom positions. The driving force for the CDW in two-dimensional metals may be described in several ways. One explanation is based on Peierls idea of electronic instability in one-dimensional metals~\cite{PeierlsRE1955}. When large parallel regions of Fermi surfaces can be connected by a single wave vector, which is referred to as nesting, the static generalized susceptibility diverges. This leads to the effective screening of phonons by conduction electrons at the wave vector~\cite{KohnW1959}, which induces a static periodic lattice distortion. One major theoretical issue concerns the fragility of the Peierls instability against small deviations from perfect nesting conditions~\cite{JohannesMD2008}. For the quasi-two-dimensional transition metal dichalcogenides~\cite{WithersRL1986}, band structure calculations did not show strong nesting~\cite{WexlerG1976, WilsonJA1977}. Another explanation is given in terms of electronic instability due to a connection between relatively small areas at the saddle points of Fermi surfaces~\cite{RiceTM1975}. The third explanation is that lattice transition is not a secondary effect, and CDW is driven by the momentum-dependent electron-phonon coupling that connects electronic and lattice instability~\cite{DoranNJ1978, VarmaC1983, JohannesMD2008}. Many experimental techniques are used to explore the origin of the CDW in two-dimensional metals, such as angle-resolved photoemission spectroscopy (ARPES)~\cite{LiuR1998}, scanning tunneling microscopy~\cite{WangC1990}, X-ray diffraction (XRD) under a magnetic field~\cite{DuCH2000, ChangJ2012}, resonant X-ray diffraction~\cite{GhiriG2012, CominR2015} and combining ARPES and XRD~\cite{RitschelT2015}. However, the origins of CDW have yet to be adequately verified. In addition, the relationship between CDW and superconductivity remains under debate. Although the CDW and superconductivity are seemingly competitive if they occur in the same part of Fermi surface, it was demonstrated that the CDW order in 2$H$-NbSe$_2$ cooperates with the superconductivity at the same wave vector due to electron-phonon coupling~\cite{KissT2007}. One of the reasons why the mechanism for CDW and the relation to superconductivity are inconclusive is the limited materials that exhibit CDW and superconductivity. Thus, the identification of a new material that exhibits CDW and superconductivity could provide a better understanding.

The wide variety of {\cabe}-type superconductors and their derivatives that have been recently discovered will be suitable for the study of CDW in two-dimensional systems and the relation to superconductivity. The coexistence of CDW and superconductivity was observed in the Pt-based 122 compounds, such as {\sras}~\cite{KudoK2010}, BaPt$_2$As$_2$~\cite{JiangWB2015}, and LaPt$_2$Si$_2$~\cite{SheltonRN1984}. For example, {\sras} exhibits superconductivity at the critical temperature ${\tc} = 5.2$~K~\cite{KudoK2010, XuX2013}, and the CDW order at the transition temperature ${\tz} = 470$~K~\cite{ImreA2007, FangAF2012, WangL2014}. The high-temperature (HT) phase of {\sras} has a tetragonal {\cabe}-type structure~\cite{ImreA2007, FangAF2012}. The As-Pt-As and Pt-As-Pt block layers in the HT phase of {\sras} are stacked alternately along the c-direction, and Sr ions are intercalated between them. {\sras} has a phase transition at {\tz} to an orthorhombic structure that has incommensurate modulation with wave vectors $q_1 = 0.62a^{\star}$ on the $a-b$ plane and $q_2 = 0.23a^{\star}$ on the $a-c$ plane~\cite{ImreA2007, WangL2014}. Imre {\etal} ascribed the CDW order to Peierls instability using band structure calculations~\cite{ImreA2007}. Kim {\etal} ascribed this to momentum-dependent electron-phonon coupling using band structure calculations~\cite{KimS2015}; a weak nesting feature and evidence of a sizable electron-phonon interaction were found. However, the origin is still under debate. In {\sr}, the coexistence of CDW and superconductivity is expected, as with {\sras}, while the coexistence of both phenomena is still unclear. It was reported that {\sr} exhibits superconductivity at ${\tc} = 2.1$~K. Electrical resistivity measurements indicated a first-order phase transition at ${\tz} \sim 270$~K~\cite{ImaiM2013}, which implies a CDW transition. The structure was first reported by Imre {\etal}~\cite{ImreA2007} to be a {\cabe}-type structure at room temperature. In contrast, the XRD pattern reported by Imai {\etal}~\cite{ImaiM2013} clearly demonstrated that {\sr} has not a tetragonal structure, but a lower symmetrical structure. However, the details of the structure in both phases and the magnetic properties near $T = {\tz}$ have yet to be identified. Therefore, in this study, we have investigated the crystal structure and magnetic susceptibility of both phases to clarify the phase transition at $T = {\tz}$ in {\sr}, and to provide a new example for the coexistence of CDW and superconductivity.

\section{Experimental details}
Polycrystalline {\sr} was synthesized in two steps: arc-melting and remelting of the arc-melted ingot. The sample consists mainly of the {\sr} phase, and contains only a small amount of PtSb and Sr$_x$Pt$_3$Sb$_{2-x}$ ($x = 0.4$) as impurity phases. The details have been described elsewhere~\cite{ImaiM2013}.
The crystal structure was characterized using powder XRD (RINT TTR-III, Rigaku) with Cu K$\alpha$ radiation (40~kV/300~mA). A liquid N$_2$ cryostat with a copper plate holder was used to cool the sample down to 90~K. XRD patterns were analyzed using the ``FULLPROF" software~\cite{Fullprof}.
Magnetic susceptibility measurements were conducted using a commercial superconducting quantum interference device magnetometer (MPMS, Quantum Design) with an applied field of $H = 1000$~Oe.

\section{Results}
\begin{figure}
\begin{center}
    \includegraphics[width=0.8\hsize]{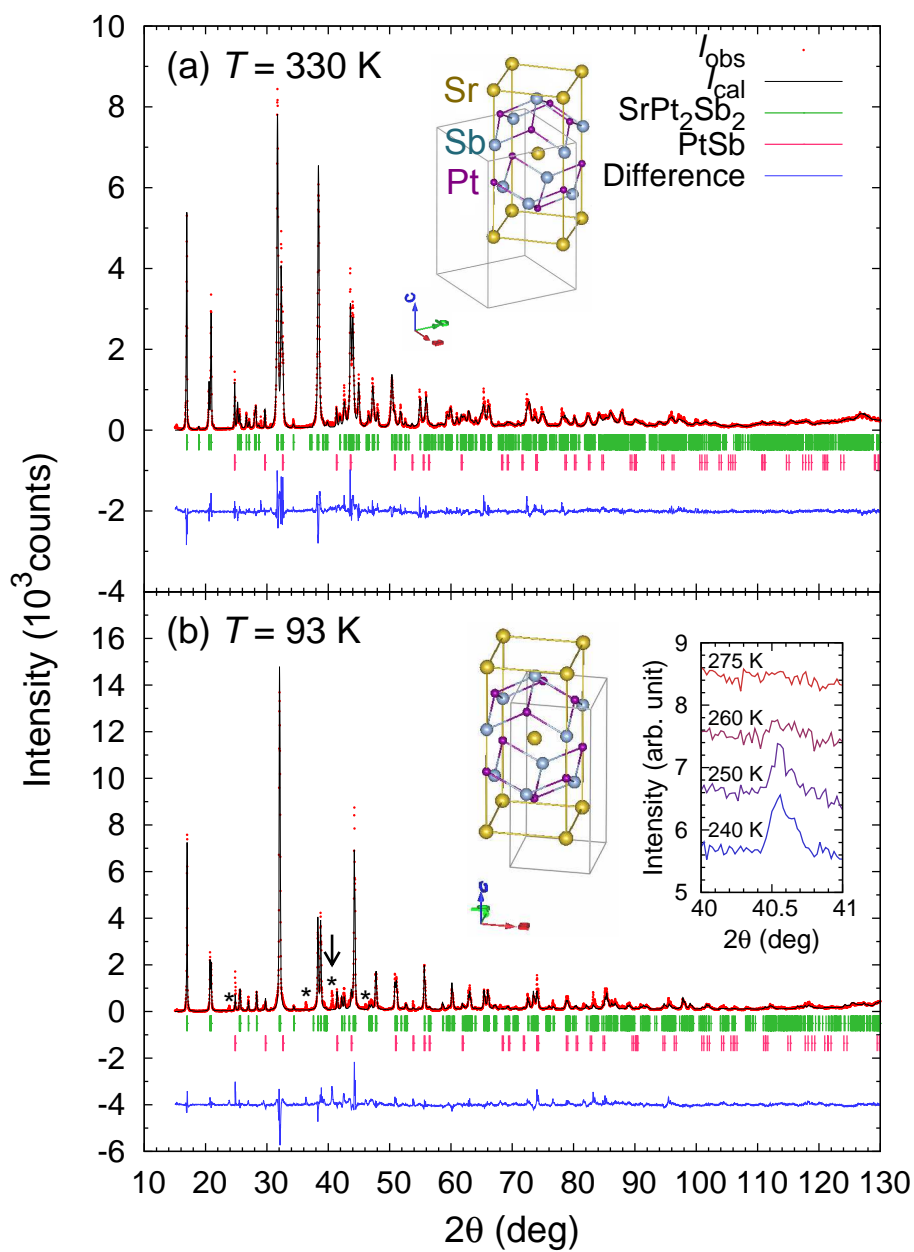}
    \caption{\label{fig1} (a)-(b) XRD patterns for the LT and HT phases measured at $T = 330$~K and 93~K, respectively. Calculated patterns are shown with black lines. The solid green and purple bars represent the Bragg peak positions for {\sr} and PtSb, respectively. Probable structures are shown in the middle insets. In (b), the peaks indicated by asterisks are those which could not be indexed with integer indices. The right inset of (b) shows the temperature dependence of the peak marked by an arrow. }
\end{center}
\end{figure}
Figure~\ref{fig1}(a) shows a powder XRD pattern for the HT phase of {\sr} measured at $T = 330$~K, in addition to the calculated pattern.
The Bragg reflections were indexed well to monoclinic symmetry with lattice parameters of $a = 6.5664(2)$~{\AA}, $b = 6.6728(2)$~{\AA}, $c = 10.4349(3)$~{\AA} and ${\beta} = 91.705(2){^\circ}$. The systematic absence of reflections suggests that the candidate space groups are $Cm, C2$ or $C2/m$. 
Assuming that the sample has the BaPt$_2$Sb$_2$-type structure~\cite{ImaiM2015}, of which the space group is $C2/m$, then the calculated pattern reproduces the observed XRD pattern with $R$ factors~\cite{Rietveld} of $R_{\rm exp} = 5.8\%$, $R_{\rm p} = 12.5\%$ and $R_{\rm wp} = 16.3\%$.
This indicates that the HT phase of {\sr} has a structure identical to BaPt$_2$Sb$_2$.

Figure~\ref{fig1}(b) shows the XRD pattern for the low-temperature (LT) phase measured at $T = 93$~K, in addition to the calculated pattern. Almost all of the peaks were successfully indexed to orthorhombic symmetry with lattice parameters of $a = 4.6387(2)$~{\AA}, $b = 4.6902(1)$~{\AA} and $c = 10.4125(2)$~{\AA}. The candidate space groups were $P2_12_12$ or $Pmmn$. 
It is interesting that the structure of the LT phase has higher symmetry than that of the HT phase.
Several small peaks marked by asterisks in figure~\ref{fig1}(b) could not be indexed with integer indices, even if triclinic symmetry was assumed.
These peaks are located at $2\theta = 23.9, 36.4, 40.6$ and $46.1{^\circ}$.
The inset of figure~\ref{fig1}(b) presents the temperature dependence of one of these peaks (denoted by the arrow). The peak emerged at {\tz}, which indicates that the peak was not a reflection from an impurity phase but a satellite reflection of the main phase. These results demonstrate that the LT phase forms a modulated structure. 
Assuming that the average structure of the LT phase is the same as that of the {\sras} LT phase~\cite{ImreA2007}, of which the space group is $Pmmn$, then the calculated pattern reproduces the observed XRD pattern, except for the satellite peaks, with $R$ factors of $R_{\rm exp} = 5.96\%$, $R_{\rm p} = 18.4\%$ and $R_{\rm wp} = 24.2\%$.
This suggests that the average structure of the LT phase is identical to that of {\sras}. The derived structures for the HT and LT phases are shown in the insets of figures~\ref{fig1}(a) and (b), respectively. 

\begin{figure}
\begin{center}
    \includegraphics[width=0.75\hsize]{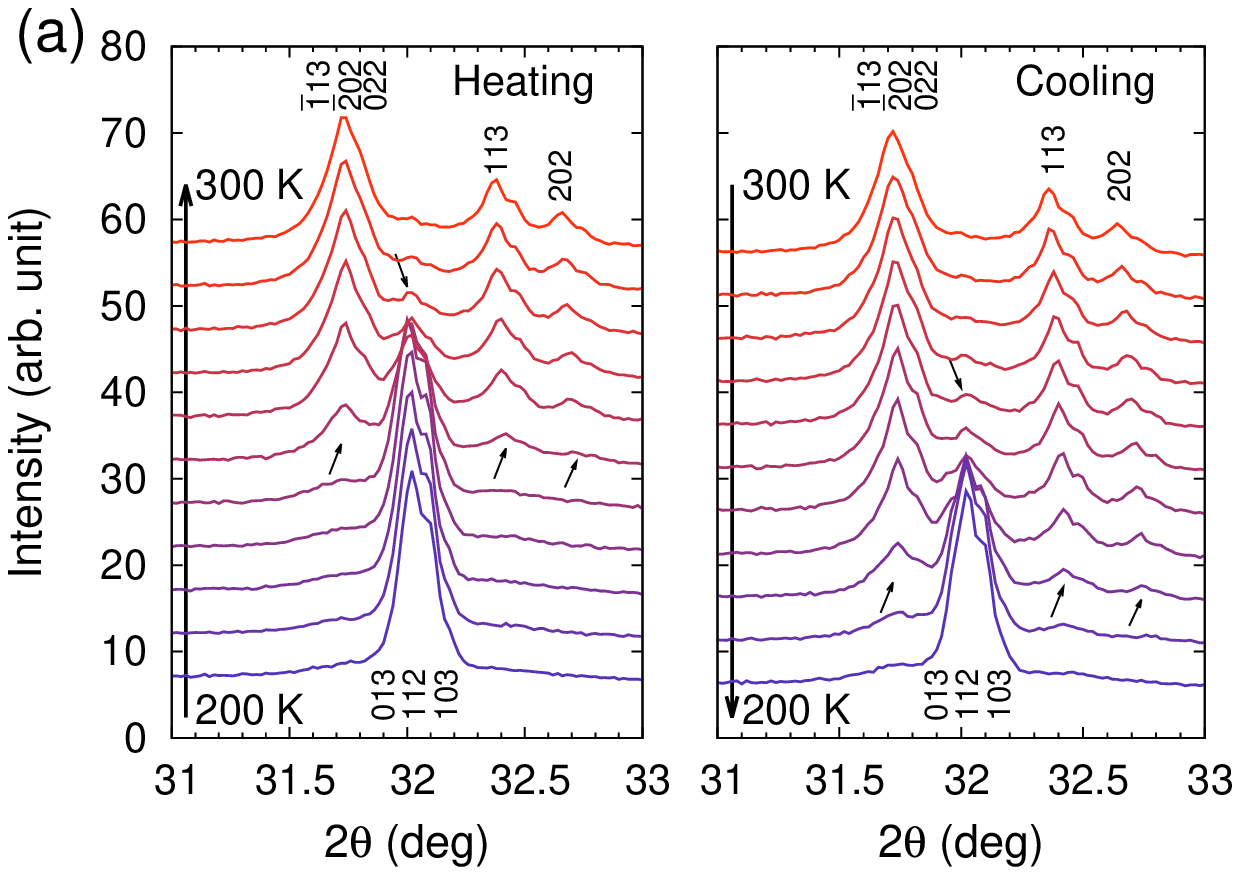}
    \includegraphics[width=0.55\hsize]{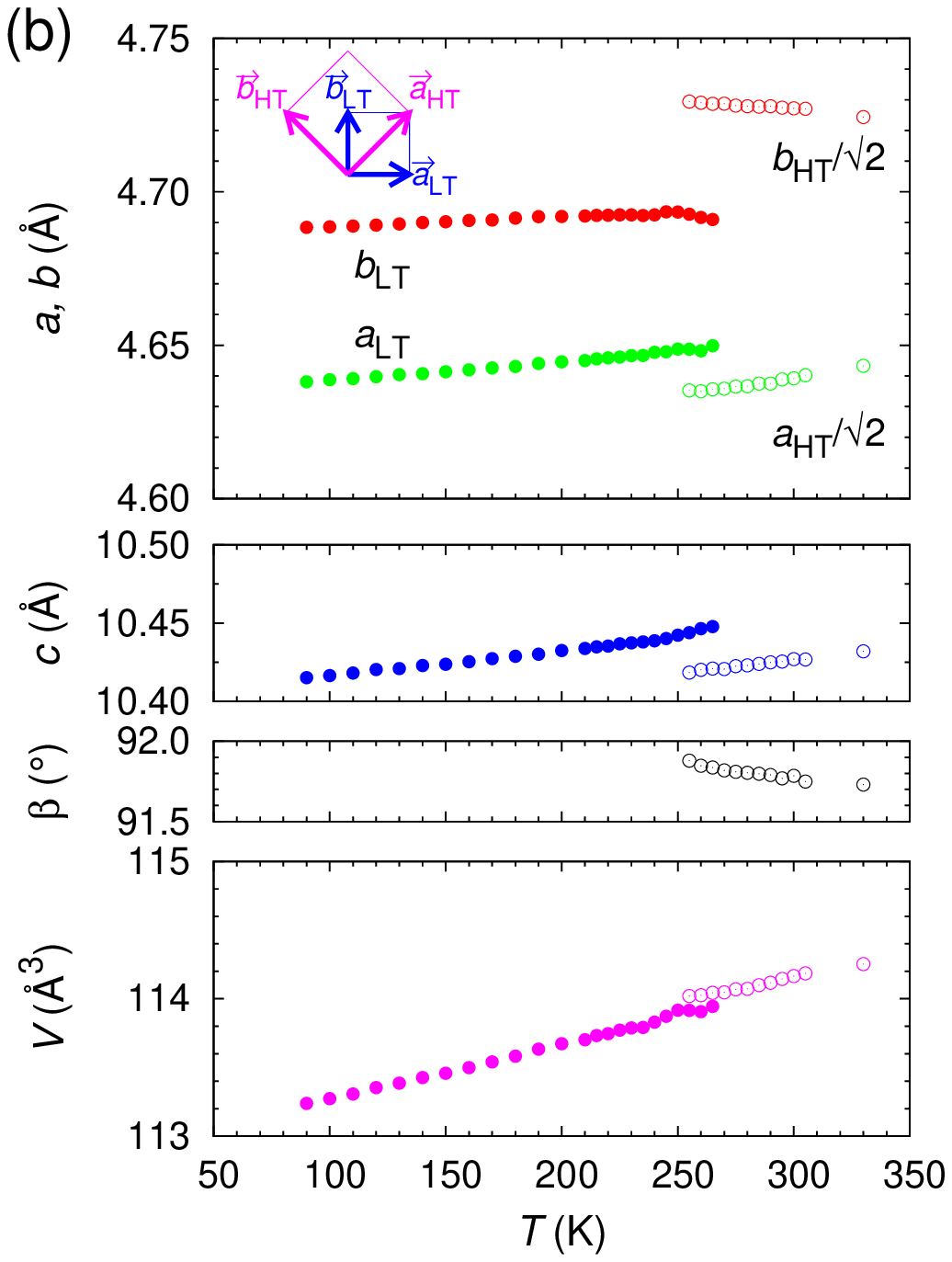}
    \vspace{30pt}
    \caption{\label{fig2} (a) XRD patterns for {\sr} around $2\theta = 32^{\circ}$ measured every 10~K from 200 to 300~K. The left and right panels show the patterns obtained by heating and cooling processes, respectively. The thin arrows represent the emergence of peaks. (b) Temperature dependence of the lattice constants ($a, b, c$ and ${\beta}$) and volume per chemical unit ($V$) for {\sr}. The solid and open circles represent the LT and HT phases, respectively. }
\end{center}
\end{figure}
Figure~\ref{fig2}(a) shows XRD patterns for {\sr} in the $2{\theta}$ range from 31 to $33^{\circ}$ every 10~K around {\tz}. The transformation of the patterns near {\tz} clearly reveals that the transition at {\tz} is accompanied by a structural transformation. The temperature region where the HT and LT phases coexist is different for the heating and cooling processes: the peaks for the HT and LT phases coexist from 250 to 280~K in the heating process, while they are present from 220 to 260~K in the cooling process. Figure~\ref{fig2}(b) shows the temperature dependence of the lattice constants and the volume per chemical formula unit, $V$, which was determined by the patterns for the heating process. $a$, $b$ and $c$ for the LT phase increase with the temperature. At the transition from the LT phase to the HT phase, $V$ increases slightly, while $c$ decreases, which indicates that the transition leads to a compression along the $c$-axis and an expansion in the $a-b$ plane. The inset of figure~\ref{fig2}(b) shows that the translation vectors for the HT phase, $a_{\rm HT}$ and $b_{\rm HT}$, are rotated by $45^{\circ}$ with respect to those of the LT phase, $a_{\rm LT}$ and $b_{\rm LT}$, and the former are approximately $\sqrt{2}$ times larger than the latter. $a/\sqrt{2}$ and $b/\sqrt{2}$ for the HT phase were plotted for comparison. In the HT phase, $a$ and $c$ increase with temperature, while $b$ decreases. The approach of $b$ to $a$ indicates a reduction of the in-plane distortion. In addition, $\beta$ approaches $90^{\circ}$ with increasing temperature, which indicates a reduction of the monoclinic distortion. These results imply the existence of another structural transition back to an orthorhombic or tetragonal structure at a higher temperature. 
\begin{table}
\caption{\label{tab1} Thermal expansion coefficients for the LT and HT phases.}
\begin{indented}
\lineup
\item[]\begin{tabular}{@{}llllll}
\br
 & $a$ & $b$ & $c$ & $\beta$ & $V$\\
 & ($10^{-6}$/K) & ($10^{-6}$/K) & ($10^{-6}$/K) & ($10^{-6}$/K) & ($10^{-6}$/K)\\
\mr
LT & 14 &\05.8 & 15 & --- & 35\\
HT & 27 & \-16 & 18 & \-16 & 30\\
\br
\end{tabular}
\end{indented}
\end{table}
Table~\ref{tab1} summarizes the thermal expansion coefficients for the LT and HT phases. These values are comparable to those of typical intermetallic compounds.

\begin{figure}
\begin{center}
    \includegraphics[width=0.75\hsize]{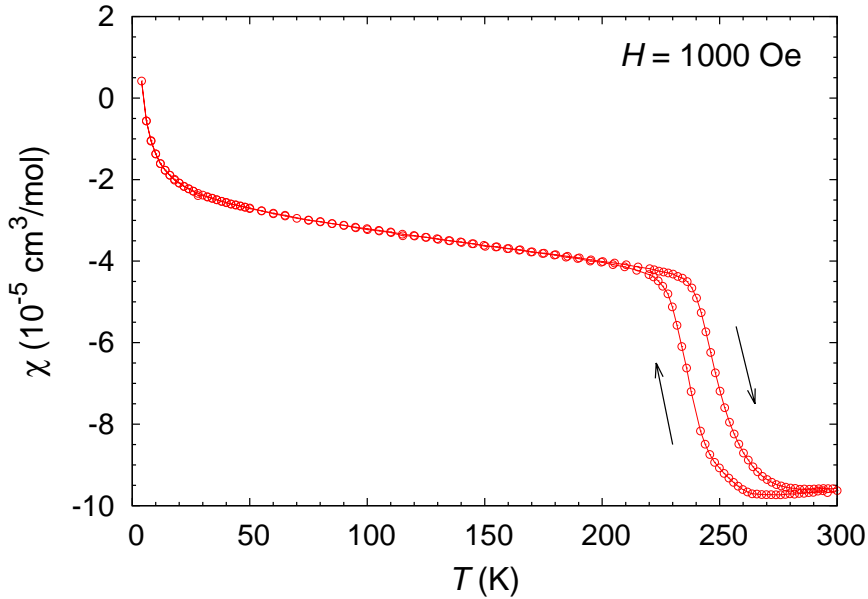}
    \vspace{30pt}
    \caption{\label{fig3} Temperature dependence of the magnetic susceptibility ($\chi$) for {\sr}. The arrows denote heating and cooling processes.}
\end{center}
\end{figure}
Figure~\ref{fig3} presents the temperature dependence of the magnetic susceptibility for {\sr}. The magnetic susceptibility at 300~K was approximately $-1\times10^{-4}$~cm$^3$/mol. The weak negative susceptibility reflects the contribution of Larmor diamagnetism and Pauli paramagnetism. A clear up-turn was observed at {\tz} with decreasing temperature, which indicates an increase in the electron density of states (DOS) at the Fermi energy. This result clearly demonstrates that the transition at {\tz} is not a magnetic transition. In addition, thermal hysteresis was observed. The small up-turn at low temperatures can be attributed to paramagnetic impurities. 

At the phase transition, both the lattice parameters and the volume change discontinuously, which indicates that this structural phase transition is a first-order phase transition. This is consistent with the observation of thermal hysteresis in the XRD, magnetic susceptibility, and previously reported electrical resistivity measurements~\cite{ImaiM2013}. The observation of the first-order phase transition is also consistent with the crystal structures of the LT and HT phases because space group theory forbids a second-order phase transition between the space group of the LT phase ($Pmmm$) and that of the HT phase ($C2/m$)~\cite{IntTab2002}.

\section{Discussion}
\begin{table}
\caption{\label{tab2} Transition temperatures, {\tc} and {\tz}, and structures of the LT and HT phases for {\cabe}-type and derivative superconductors.}
\begin{indented}
\lineup
\item[]\begin{tabular}{@{}llllll}
\br
 & {\tc} (K) & {\tz} (K) & LT Struct. & HT Struct. & References \cr
\mr
SrPt$_2$As$_2$ & 5.2 & 470 & Orth. + mod. & Tetra. & \cite{KudoK2010, ImreA2007, FangAF2012, WangL2014}\cr 
BaPt$_2$As$_2$ & 1.33, 1.67 & 275 & Orth. & Tetra. & \cite{JiangWB2015}\cr 
SrPt$_2$Sb$_2$ & 2.1 & 270 & Orth. + mod. & Mono. & \cite{ImaiM2013}, This study\cr 
BaPt$_2$Sb$_2$ & 1.8 & --- & --- & Mono. & \cite{ImaiM2015}\cr 
LaPt$_2$Si$_2$ & 1.31--1.42 & 112 & Orth. + mod. & Tetra. & \cite{SheltonRN1984, NaganoY2013}\cr 
YPt$_2$Si$_2$ & 1.5 & --- & --- & Tetra. & \cite{NaganoY2013}\cr 
LaPd$_2$Sb$_2$ & 1.4 & --- & --- & Tetra. & \cite{GanesanpottiS2014}\cr 
\br
\end{tabular}
\item[] Tetra.: tetragonal; Orth.: orthorhombic; Mono.: monoclinic; mod.: modulation.
\end{indented}
\end{table}

The structural and superconducting transition temperatures for the {\cabe}-type and derivative superconductors are listed in table~\ref{tab2}. The coexistence of the CDW and superconductivity has been proposed for some of these superconductors: {\sras}, BaPt$_2$As$_2$ and LaPt$_2$Si$_2$. In these compounds, orthorhombic distortion from the {\cabe}-type tetragonal structure was observed and associated with the CDW order~\cite{KudoK2010, ImreA2007, FangAF2012, WangL2014, JiangWB2015, SheltonRN1984, NaganoY2013}. {\sr} showed a transition from the monoclinic to a modulated orthorhombic structure. The plausible explanation for the phase transition at {\tz} in metallic {\sr} will be the CDW order because of the modulated structure and the absence of magnetic order in the LT phase. {\sr} can thus be a new example for the coexistence of the CDW and superconductivity. On the other hand, other mechanisms could not be ruled out. The crystal system at room temperature is monoclinic for the compounds consisting of heavy ions: {\sr} and BaPt$_2$Sb$_2$. Orbitals may participate in the structural transition in some way.

The magnetic susceptibility and electrical resistivity of these compounds reveal different changes at {\tz}. An increase in the Pauli paramagnetism component of the magnetic susceptibility is caused by an increase in the DOS at the Fermi energy. An increase in the electrical resistivity is caused by a decrease in the DOS at the Fermi energy and/or an increase in electron scattering by phonons. The magnetic susceptibility remains unchanged with decreasing temperature for BaPt$_2$As$_2$~\cite{JiangWB2015}, whereas it decreases for LaPt$_2$Si$_2$~\cite{NaganoY2013} and increases for {\sr}. To the best of our knowledge, the magnetic susceptibility of {\sras} near {\tz} has not been reported to date. The electrical resistivity decreases with decreasing temperature for {\sras}~\cite{FangAF2012, WangL2014}, but increases for BaPt$_2$As$_2$~\cite{JiangWB2015}, LaPt$_2$Si$_2$~\cite{NaganoY2013} and {\sr}. The CDW order causes opening of the energy gap at the nesting position. If large parts of the Fermi surfaces are nested, as supposed by the Peierls instability scenario, then a decrease in the magnetic susceptibility and an increase in the electrical resistivity with decreasing temperature are expected due to the large decrease in the DOS. If it is assumed that the CDW has the same origin in these compounds, then the variety of changes in the magnetic susceptibility and electrical resistivity indicates that the nesting is limited to small parts of the Fermi surfaces, and that the changes are mainly dependent on the reconstruction of the electronic structure by the structural change. This is consistent with optical conductivity results in {\sras}~\cite{FangAF2012} and the quasi-nesting features obtained by band structure calculations in {\sras}~\cite{KimS2015}. The negative curvature of the electrical resistivity temperature dependence (${\rm d}^2\rho/{\rm d}T^2 < 0$)~\cite{FangAF2012, WangL2014, JiangWB2015, NaganoY2013, ImaiM2013} indicates strong electron-phonon interaction~\cite{WoodardDW1964, FiskZ1976, AllenPB1978}. These results suggest that the CDW instability originates from momentum-dependent electron-phonon coupling. Further microscopic studies on the electronic structures and phonon softening will be necessary using ARPES and neutron scattering, respectively, to clarify the origin of the CDW. 

Among the compounds listed in table~\ref{tab2}, {\sras} has particularly large {\tc} and {\tz}, which suggests that the superconductivity is enhanced by the CDW. However, the other compounds show similar {\tc} between 1.3 and 2.1~K, regardless of the CDW order, which indicates that there is no close correlation between superconductivity and the CDW. In addition, {\sras} and other {\cabe}-type materials have similar electronic structures, as determined from band calculations~\cite{NekrasovIA2010, SheinIR2011, HaseI2013, ImaiM2015, KimS2015}. The main contributors to the DOS at the Fermi energy are the $d$ states of Pt ions in the As-Pt-As layer of {\sras}~\cite{NekrasovIA2010, SheinIR2011, KimS2015}. Similar trends have been reported for BaPd$_2$Sb$_2$~\cite{HaseI2013} and BaPt$_2$Sb$_2$~\cite{ImaiM2015}. Therefore, this may be a general trend for the {\cabe}-type materials~\cite{ZhengC1986}. The large difference of {\tc} and {\tz} between {\sras} and the other compounds may thus be attributed to electron-phonon interaction. Additional studies on electron-phonon interactions could thus reveal the relation between superconductivity and the CDW of {\cabe}-type and derivative compounds.

\section{Summary}
The first-order transition of {\sr} around ${\tz} = 270$~K was investigated using XRD and magnetic susceptibility measurements. 
In the LT phase, the structure was modulated and no magnetic order was observed. 
These results suggest that the transition at {\tz} originates from the CDW order. The magnetic susceptibility increased with decreasing temperature at $T \sim \tz$, where the change was different from those observed for BaPt$_2$As$_2$~\cite{JiangWB2015} and LaPt$_2$Si$_2$~\cite{NaganoY2013}. This indicates that the CDW gap is limited to small parts of the Fermi surfaces in {\sr}. In addition, it is unique that the average structure of the LT phase has higher symmetry than that of the HT phase. To reveal the interplay of the CDW and superconductivity in {\cabe}-type and derivative compounds, further research on the electron-phonon interaction will be required.

\ack
The authors are indebted to CROSS-Tokai for providing the MPMS at their user laboratories. Part of this work was supported by the Japan Society for the Promotion of Science (JSPS) through the Funding Program for World-Leading Innovative R\&D on Science and Technology (FIRST Program).

\section*{References}
\bibliography{SrPt2Sb2}
\end{document}